\documentclass[11pt]{article} 

\usepackage[colorlinks=true, allcolors=blue]{hyperref}

\usepackage[utf8]{inputenc} 

\usepackage{geometry} 
\usepackage{graphicx} 

\usepackage[round]{natbib}

\usepackage{booktabs} 
\usepackage{array} 
\usepackage{paralist} 
\usepackage{verbatim} 
\usepackage{subfig} 

\usepackage{amsmath}
\usepackage{amsfonts}
\usepackage{amssymb}
\usepackage{mathrsfs}

\usepackage{multirow}

\allowdisplaybreaks

\newcommand{\RRR}{\mathbb{R}}
\newcommand{\EEE}{\mathbb{E}}
\newcommand{\PPP}{\mathbb{P}}
\newcommand{\dd}{\mathrm{d}}

\newcommand{\rA}{\mathrm{A}}
\newcommand{\rB}{\mathrm{B}}

\newcommand{\cN}{\mathcal{N}}

\newcommand{\Bi}{\mathrm{B}}

\newcommand{\Hu}{\mathrm{H}}
\newcommand{\Qu}{\mathrm{Q}}

\newcommand{\CSI}{\mathrm{CSI}}

\usepackage{fancyhdr} 
\usepackage{sectsty}
\allsectionsfont{\sffamily\mdseries\upshape} 

\usepackage[nottoc,notlof,notlot]{tocbibind} 
\usepackage[titles,subfigure]{tocloft} 

\title{A scoring framework for tiered warnings and multicategorical forecasts based on fixed risk measures}

\author{Robert Taggart, Nicholas Loveday, Deryn Griffiths\\Australian Bureau of Meteorology\\robert.taggart@bom.gov.au}

\begin{document}

\maketitle

\begin{abstract}
\noindent The use of tiered warnings and multicategorical forecasts are ubiquitous in meteorological operations. Here, a flexible family of scoring functions is presented for evaluating the performance of ordered multicategorical forecasts. Each score has a risk parameter $\alpha$, selected for the specific use case, so that it is consistent with a forecast directive based on the fixed threshold probability $1-\alpha$ (equivalently, a fixed $\alpha$-quantile mapping). Each score also has use-case specific weights so that forecasters who accurately discriminate between categorical thresholds are rewarded in proportion to the weight for that threshold. A variation is presented where the penalty assigned to near misses or close false alarms is discounted, which again is consistent with directives based on fixed risk measures. The scores presented provide an alternative to many performance measures currently in use, whose optimal threshold probabilities for forecasting an event typically vary with each forecast case, and in the case of equitable scores are based around sample base rates rather than risk measures suitable for users.
\vspace{10pt}

\end{abstract}

\section{Introduction}

A broad transdisciplinary consensus has emerged over the last two of decades that forecasts ought to be probabilistic in nature, taking the form of a predictive distribution over possible future outcomes \citep{ehm2016quantiles, gneiting2014probabilistic}. However, in certain settings categorical forecasts are still very useful, particularly when there is a need for simplicity of communication or to trigger clear actions. Examples include public weather warnings, alerting services designed to prompt specific protective action for a commercial venture, and forecasts displayed as graphical icons.

Ideally, any categorical forecast service will be designed so that following the forecast directive, optimising the performance score and maximising benefits for the user are \textit{consistent} \citep{murphy1985forecast, murphy1993good, gneiting2011making}. In this context, a forecast directive is a rule to convert a probabilistic forecast into a categorical forecast. The consistent scoring function can be used to track trends in forecast performance, guide forecast system improvement and rank competing forecast systems, so that decisions made on such evaluations are expected to benefit the user. When the target user group is heterogeneous, maximising user benefit will not be possible. Nonetheless, in the context of public warning services for natural hazards, issues such as warning fatigue, false alarm intolerance and the cost of missed events have received considerable attention (e.g. \cite{gutter2018severe, potter2018influence, mackie2014warning, hoekstra2011preliminary}). This body of literature, along with stakeholder engagement, should provide guidance for creating a suitable forecast directive for the particular hazard.

Performance measures for categorical forecasts in the meteorological literature typically have properties that are undesirable for many applications. For example, \textit{equitable} scores for multicategorical forecasts, such as those of \cite{gandin1992equitable}, of which the \cite{gerrity1992note} score is perhaps most popular, are optimised by routinely adjusting the threshold probability used to convert a probabilistic forecast into a categorical one. Moreover, these variable threshold probabilities are usually near climatological or sample base rates, which could lead to a proliferation of false alarms when warning for rare events. The Murphy and Gandin scores also penalise false alarms and misses equally, even though the costs of such errors to the user are rarely equal. In Section~\ref{s:critique} we show that, depending on the use case, these and other performance measures could potentially lead to undesirable or even perverse service outcomes if forecasters took score optimisation seriously. Nevertheless, we found many examples in the literature where the suitability of the measure for the problem at hand was not adequately discussed, possibly because the implications of score optimisation were not well understood.

To provide alternatives, in Section~\ref{s:framework} we present a family of scoring functions for ordered categorical forecasts that have flexibility for a broad range of applications, particularly public forecast services, and are consistent with directives based on fixed risk thresholds. A scoring function is a rule that assigns a penalty for each particular forecast case when it is compared to the corresponding observation. In discussing ordered categorical forecast services, we are assuming that forecasts are for some unknown real-valued quantity, such as accumulated precipitation, and that these must be issued as a categorical forecast rather than a real-valued forecast or a predictive distribution.

Within this framework, those designing a multicategorical forecast service specify
\begin{enumerate}
\item[(i)] category thresholds that delineate the categories, 
\item[(ii)] weights that specify the relative importance of forecasting on the correct side of each category threshold, and
\item[(iii)] the relative cost $\alpha/(1-\alpha)$ of a miss to a false alarm, where $0<\alpha<1$.
\end{enumerate}
In this setup, the scoring function can be expressed as a scoring matrix, which specifies the penalty for each entry in the contingency table. The scoring function is consistent with the directive `Forecast a category which contains an $\alpha$-quantile of the predictive distribution.' In the dichotomous case this directive is equivalent to `Forecast the event if and only if $\PPP(\mathrm{event})>1-\alpha$.' The weights given by (ii) are used for scoring, so that forecasters who can accurately discriminate between events and nonevents at thresholds with higher weights are appropriately rewarded. We show that the scoring matrix has a natural interpretation in terms of the simple, classical cost--loss decision model \citep{thompson1952operational, murphy1977value, richardson2000skill}.

A natural extension of this framework is presented in Section~\ref{s:variations}, with the additional specification of
\begin{enumerate}
\item[(iv)] a discounting distance parameter $a$, where $0\leq a\leq\infty$, such that the cost of misses and false alarms are discounted whenever the observation is within distance $a$ of the forecast category.
\end{enumerate}
When $a=0$ no discounting occurs and the setup in the previous paragraph is obtained. When $a>0$, the framework gives a scoring function that discounts the cost of near misses and close false alarms. \cite{barnes2007false} argued that this was a desirable property in certain contexts, without providing such a scoring function. In this case, the consistent directive is expressed in terms of a \textit{Huber quantile} \citep{taggart2022point} rather than a quantile of the predictive distribution. In the limiting case $a=\infty$, where forecast errors are penalised in proportion to the distance from the forecast category to the real-valued observation, the directive is expressed in terms of an \textit{expectile} of a predictive distribution. Expectiles, of which the expected value is most widely used, have recently attracted interest in finance as measures of risk \citep{bellini2017risk}. The relationship between quantiles, Huber quantiles and expectiles is summarised in Section~\ref{ss:discount variation} and illustrated in Figure~\ref{fig:huber}. This paper demonstrates that these statistical functionals have applications in meteorology as well as finance.

A special case of this framework ($\alpha=1/2$ and $a=\infty$) covers the situation where each forecast category indicates the likelihood of an event, such as there being a `low`, `medium` or `high` chance of lightning. The corresponding consistent scoring matrix is presented at the end of Section~\ref{s:variations}.

Since score optimisation is consistent with directives based on fixed risk measures, we refer to this framework as the FIxed Risk Multicategory (FIRM) Forecast Framework and the corresponding scores as the FIRM Scores.

For public warning services, we discuss issues that may influence the choice of parameters, with particular focus on whether the threshold probability $1-\alpha$ above which one issues a warning should vary with warning lead time, and how one can estimate the implicit value of $\alpha$ in an existing service where it hasn't been specified.

The mathematics underpinning our results rests on the insights of \cite{ehm2016quantiles}, who showed that the consistent scoring functions of quantiles and expectiles are weighted averages of corresponding \textit{elementary scoring functions}, and a result of \cite{taggart2022point} who showed the same for Huber quantiles. The new scoring functions presented in this paper are linear combinations of the elementary scoring functions, adapted to the categorical context.

\section{Notation and conventions}\label{notation}

Suppose that $Y$ is some as yet unknown quantity for which one can issue a forecast, taking possible values in some interval $I \subseteq \RRR$. For example $Y$ might be the accumulated rainfall at a particular location for a specified period of time and $I=[0,\infty)$. The prediction space $I$ is partitioned into $N+1$ mutually exclusive ordered categories $(C_i)_{i=0}^N$ via \textit{category thresholds} $\{\theta_i\}_{i=1}^N\subset I$ that satisfy $\theta_1<\theta_2<\ldots<\theta_N$. That is, $Y$ lies in category $C_0$ if $Y\leq\theta_1$, in the category $C_i$ for $1\leq i<N$ if $\theta_i<Y\leq\theta_{i+1}$, and in the category $C_N$ if $Y>\theta_N$. Thus we assume that each category includes the right endpoint of its defining interval, noting that the theory is easily adapted for those who prefer the opposite convention.

A predictive distribution issued by some forecaster for the quantity $Y$ will be denoted by $F$ and identified with its cumulative density function (CDF). Hence $\PPP(Y\leq y) = F(y)$ for each possible outcome $y$ in $I$. From $F$ one obtains the forecast probability $p_i$ that $Y$ lies in the category $C_i$, namely
\[
p_i = 
\begin{cases}
F(\theta_1), & i = 0,\\
F(\theta_{i+1}) - F(\theta_i), & 1\leq i<N,\\
1-F(\theta_N), & i = N.
\end{cases}
\]
If $0<\alpha<1$ then the set of $\alpha$-quantiles of the predictive distribution of $F$ will be denoted by $\Qu^\alpha(F)$, noting that in meteorological applications $\Qu^\alpha(F)$ is typically a singleton. The CDF of the standard normal distribution $\cN(0,1)$ will be denoted by $\Phi$, and its probability density function (PDF) by $\phi$.

In the context of tiered warnings, we adopt the convention that higher values of $Y$ represent more hazardous conditions. Thus $C_0$ can be interpreted as the category of having no warning, while $C_i$ represents more hazardous warning categories as nonzero $i$ increases. In some practical cases, such as warning for extremely cold conditions, the reverse convention is more applicable and the theory can be adapted appropriately.

For a fixed set of forecast cases, the contingency table $(c_{ij})_{i,j=0}^N$ is a complete summary of the joint distribution of categorical forecasts and corresponding categorical observations. Here $c_{ij}$ is the number of cases for which category $C_i$ was forecast and $C_j$ observed. See the two 3 by 3 arrays embedded in  Table~\ref{tab:rainfall contingency} for an example. In the binary case, where category $C_0$ is interpreted as a nonevent and $C_1$ as an event, we denote $c_{00}$ by $c$ (the number of correct negatives), $c_{01}$ by $m$ (the number of misses), $c_{10}$ by $f$ (the number of false alarms) and $c_{11}$ by $h$ (the number of hits). Note that, to maintain consistency with higher category cases, the contingency table for dichotomous forecasts is in reverse order to the usual convention.

\section{Service implications of optimising performance measures}\label{s:critique}

Many commonly used performance measures for categorical forecasts could lead to undesirable or possibly perverse service outcomes if forecasters were to take score optimisation seriously. Some common issues that arise will be illustrated by considering the probability of detection (POD) and false alarm ratio (FAR), the Critical Success Index (CSI), the Gerrity score and Extremal Dependence Score (EDS). These, and other measures, have their uses, but their alignment with service outcomes should be carefully assessed prior to their employment \citep{mason2003binary}.

\subsection{POD and FAR}

For dichotomous forecasts, POD and FAR, defined by
\[\mathrm{POD} = \frac{h}{h+m} \quad\text{and}\quad \mathrm{FAR}= \frac{f}{h+f},\]
are often used in tandem to report on the accuracy of a warning system \citep{karstens2015evaluation, brooks2018long, stumpf2015probabilistic}. POD is optimised by always warning and FAR by never warning, so together they don't constitute a clear forecast directive, and in general POD and FAR cannot be used to rank competing warning systems.

There have been various attempts by those using POD and FAR to provide greater service clarity. For example, in evaluating its fire weather warnings, the Australian Bureau of Meteorology (BoM) specifies an annual target that $\mathrm{POD} \geq 0.7$ and $\mathrm{FAR} \leq 0.4$ in each of its geographic reporting regions. If an idealised forecast system issues perfectly calibrated Gaussian predictive distributions, then Appendix~\ref{a:pod and far} shows that a viable initial strategy for meeting BoM targets is to warn if $p_1\gtrsim0.3$, noting that the optimal threshold probability for warning depends on the sharpness of the predictive distributions and the observed base rate. However, if towards the end of the annual reporting period the current POD lies comfortably above 0.7 while $\mathrm{FAR} > 0.4$, then the best strategy for a forecaster seeking to meet targets would be to warn only if the event were a near certainty in an attempt to reduce FAR. Thus meeting performance targets, if taken seriously, could result in highly undesirable outcomes. Note also that for accurate forecast systems, there is little incentive to improve categorical predictive performance since the target will be met with very high probability for a wide range of warning decision strategies. Finally, stronger predictive performance is required to meet targets in geographic regions with lower base rates.

\subsection{CSI}

The CSI for a set of dichotomous forecasts is defined by $\CSI = h/(h+m+f)$ and dates back to \cite{gilbert1884finley}. It is widely used for dichotomous forecasts of infrequent or rare events \citep{karstens2015evaluation, skinner2018object, cintineo2020noaa, stumpf2015probabilistic} since it isn't dominated by the large number of correct negatives relative to other outcomes. A forecaster's expected CSI can be optimised by forecasting $C_1$ if and only if $p_1 > h/(2h+m+f)$
\citep{mason1989dependence}, where $h$, $m$ and $f$ are entries of the contingency table for cases thus far. Hence the optimal threshold probability adjusts according to forecast performance, with less skilful forecasters warning when there is a lower chance of an event compared to more skilful forecasters, and all forecasters warning when $p_1\geq 0.5$. As discussed in Section~\ref{ss:practical}, there are good reasons why a public warning service might be designed to warn at a higher level of confidence if issuing a warning earlier than the standard lead time. Optimising CSI works against this, since longer lead time forecasts typically have less skill.

\subsection{Equitable scores}

The meteorological literature on multicategorical forecasts has often proffered \textit{equitability} as a desirable property for a scoring rule (e.g. \cite{livezey2003categorical}). A score is equitable if all constant forecasts and random forecasts receive the same expected score. The family of \cite{gandin1992equitable} scores for multicategorical forecasts are constructed so they are equitable, penalise under- and over-prediction equally (\textit{symmetry}), reward correct categorical forecasts more than incorrect forecasts, and penalise larger categorical errors more heavily. The \cite{gerrity1992note} score and LEPSCAT \citep{potts1996revised} are members of this family. The Gerrity score is Livezey's leading recommendation and has been used, for example, by \cite{bannister2021pilot} and \cite{kubo2017verification} for tiered warning verification. In the 2-category case, the Gerrity score is identical to Peirce's skill score \citep{peirce1884numerical}.

We give four reasons why the Gandin and Murphy scores are unsuitable for a wide variety of applications, including many warning services. This is primarily due to the properties of equitability and symmetry.

First, the cost of false alarms and misses to users of a warning service are rarely equal.

Second, equitability ensures that the rewards for forecasting rare events are sufficiently high that forecasting the event will be worthwhile even if the likelihood of it occurring is small. These scores `do not reward conservatism' \citep[p.~84]{livezey2003categorical}, primarily because incorrect forecasts of less likely categories are penalised relatively lightly. For example, the strategy that optimises the expected Gerrity score in the dichotomous case is to warn if and only if $p_1>r_1$, where $r_1$ is the sample base rate of the event. If the forecaster estimates that $r_1<0.01$, then warning when the probability of occurrence exceeds $1\%$ is a worthwhile strategy for score optimisation, even if it leads to a proliferation of false alarms that erodes public trust in the service. A related issue is that entries of the scoring matrices include reciprocals of sample base rates, so that sampling variability results in score instability if one category is rarely observed.

Third, in higher category cases, the rule for converting a predictive distribution into a categorical forecast is not transparent. For example,  a routine if somewhat tedious calculation shows that forecasting the highest category $C_2$ for the 3-category Gerrity score is optimal if and only if
\[
p_2 > \max\left\{
\frac{(v_0^{-1} + v_2 + 2)p_0 + (v_2-v_0)p_1}{v_0 + v_2^{-1} + 2}, v_2(p_0+p_1)
\right\},
\]
where $r_i$ is the sample base rate for category $C_i$ and $v_i = r_i/(1-r_i)$ is a sample odds ratio. Since each $r_i$ also needs to be forecast, there is no clear mapping from the forecaster's predictive distribution for a particular forecast case to the optimal categorical forecast. Nor, in the case of public warnings, would it be easy to communicate service implications to key stakeholders.

Fourth, optimal rules for converting predictive distributions into categorical forecasts require on-going re-estimation of final sample base rates, using (say) a mixture of climatology and observed occurrences, which results in shifting optimal threshold probabilities. For example, in the 2-category case \cite{mason2003binary} states that the optimal strategy is to warn if and only if $p_1>(h+m+1)/(n+2)$, where $n$ is the number of forecast cases and where it is assumed that the forecaster has no climatological knowledge. A modification to this strategy that makes use of prior climatological knowledge is achievable using Bayesian techniques (c.f. \cite{garthwaite1995statistical}, Example~6.1). In either case, the optimal warning threshold probability changes with every forecast case. A forecaster may find themselves initially warning when the risk of an event exceeds 5\%, and later warning when the risk exceeds 2\%. For a department that must spend its asset protection budget each financial year or risk a funding cut, regularly adjusting threshold probabilities is warranted. But such properties should be one of choice rather than an unintended consequence of selecting an `off the shelf' performance measure. 

This fourth property is a direct consequence of equitability. It can be addressed by using a scoring matrix constructed from base rates of a fixed past climatological reference period, though the score will no longer be truly equitable. However, this adjustment will not address the first three problems listed.

\subsection{EDS}

The EDS has recently been used to measure the performance of the German Weather Service's nowcast warning system \citep{james2018nowcastmix}. The inventors of this score write that the ``optimal threshold [probability] for the EDS is zero, and so the EDS is consistent with the rule `always forecast the event.' This rule is unlikely ever to be issued as a directive and therefore the EDS will be hedgable whenever directives are employed'' \citep[p.~705]{ferro2011extremal}. The same paper urges that ``the EDS should be calculated only after recalibrating the forecasts so that the number of forecast events equals the number of observed events.'' Given these properties, we find it difficult to see the value of the EDS for public warning performance assessment although it may be valuable in other contexts.

\subsection{Summary}

For each performance measure discussed in this section, the forecaster does not know when issuing the forecast what the penalty for a particular forecast error will be. Of these, only the Gandin and Murphy scores assign a penalty to each individual forecast case, with the penalty for each error type only clear after the distribution of observations is known. These measures appear to be designed for extracting a signal of skill from an existing contingency table when there is no information on the decision process for issuing categorical forecasts. In the next section, we introduce a family of scoring functions that are fundamentally different in nature.

\section{A new framework for ordered categorical forecasts}\label{s:framework}

\subsection{Scoring matrix, optimal forecast strategy and economic interpretation}\label{ss:basic framework}

Here we describe a score to assess ordered categorical forecasts that is consistent with directives based on fixed threshold probabilities. Those designing the categorical forecast service provide the following specifications:
\begin{itemize}
\item an increasing sequence $(\theta_i)_{i=1}^N$ of category thresholds that defines the ordered sequence $(C_i)_{i=0}^N$ of $N+1$ categories;
\item a corresponding sequence $(w_i)_{i=1}^N$ of positive weights that specifies the relative importance of a forecast and corresponding observation falling on the same side of each category threshold; and
\item a single parameter $\alpha$ from the interval $(0,1)$ such that, for every category threshold, the cost of a miss relative to the cost of a false alarm is $\alpha/(1-\alpha)$.
\end{itemize}

For example, a marine wind warning service might be based on three category thresholds $\theta_1=25\,\mathrm{kt}$, $\theta_2=34\,\mathrm{kt}$ and $\theta_3=48\,\mathrm{kt}$ to demark four categories (no warning, strong wind warning, gale warning and storm warning). If the importance of forecasting on the correct side of the highest category threshold (48 kt) is twice that of the other thresholds, then set $(w_1, w_2, w_3) = (1, 1, 2)$. Selecting $\alpha = 0.7$ implies that a miss (relative cost 0.7) is more costly than a false alarm (relative cost 0.3).

In this framework, a miss relative to the category threshold $\theta_i$ occurs when the forecast is for some category below $\theta_i$ whereas the observation $y$ satisfies $y>\theta_i$. The penalty for such a miss is $\alpha w_i$. A false alarm relative to the category threshold $\theta_i$ occurs when the forecast is for some category above $\theta_i$ whereas the observation $y$ satisfies $y\leq\theta_i$. The penalty for such a false alarm is $(1-\alpha)w_i$. Hits and correct negatives relative to $\theta_i$, where the forecast category and observed category lie on the same side of $\theta_i$, incur zero penalty. When summed across all category thresholds, this scoring system gives rise to the scoring matrix $(s_{ij})_{i,j=0}^N$, whose entries give the penalty when $C_i$ is forecast and $C_j$ observed, namely
\begin{equation}\label{eq:framework scoring matrix}
s_{ij} = 
\begin{cases}
0, & i=j,\\
\alpha\displaystyle\sum_{k=i+1}^j w_k, & i < j \\
(1-\alpha)\displaystyle\sum_{k=j+1}^i w_k, & i > j.
\end{cases}
\end{equation}
For the dichotomous and 3-category cases, the scoring matrices are
\begin{equation}\label{eq:2by2 and 3by3 scoring matrix}
\begin{pmatrix}
	0							& \alpha w_1 \\
	(1-\alpha) w_1 		& 0
\end{pmatrix}
\qquad\text{and}\qquad
\begin{pmatrix}
	0							& \alpha w_1			& \alpha(w_1+w_2) \\
	(1-\alpha) w_1 		& 0						& \alpha w_2 \\
	(1-\alpha)(w_1 + w_2) & (1-\alpha)w_2 & 0
\end{pmatrix}.
\end{equation}
The entries above the zero diagonal represent penalties for misses and those below for false alarms, while correct forecasts are not penalised. 
In the multicategory case, larger over-prediction errors receive higher penalties than smaller over-prediction errors, since a larger error is a false alarm relative to more category thresholds. A similar statement holds for under-prediction penalties.

The scoring matrix presented is consistent with the directive `Forecast any category $C_i$ which contains an $\alpha$-quantile of the predictive distribution $F$.' 
In meteorological applications, $\alpha$-quantiles and hence the choice of forecast category will typically be unique. The proof of consistency, namely that a forecaster following the directive will optimise their expected score, will be given in Section~\ref{ss:reframing and proof}. An equivalent directive reformulated in categorical terms is `Forecast the highest category for which the probability of observing that category or higher exceeds $1-\alpha$.' In the dichotomous case, the directive reduces to `Warn if and only if the forecast probability of an event exceeds $1-\alpha$.' Because of its connection with measures of risk, we refer to $\alpha$ as the \textit{risk parameter} of the scoring framework.

This scoring matrix rewards forecasters who can correctly discriminate between each threshold $\theta_i$ at the $\alpha$-quantile level. The weights $w_i$ indicate the thresholds at which discrimination is most valuable, and provides a clear signal for where to target predictive improvement. This scoring matrix has a degree of transparency that is absent for the Gerrity score, particularly as the number of categories increases.

We refer to this new framework as the FIxed Risk Multicategory (FIRM) framework, because optimal forecasting strategies are consistent with forecast directives based on the fixed threshold probability $1-\alpha$, or equivalently on the $\alpha$-quantile as a measure of risk for fixed $\alpha$. The framework presented in this subsection is denoted by
\[
\mathrm{FIRM}\left((\theta_i)_{i=1}^N, (w_i)_{i=1}^N, \alpha, 0\right),
\]
where the first three parameters specify the category thresholds $\theta_i$, corresponding weights $w_i$ and risk parameter $\alpha$. The \textit{raison d'\^{e}tre} of the final parameter, called the \textit{discounting distance parameter} and here taking the value $0$, will become apparent in Section~\ref{s:variations} where an extension of the framework is presented.

The FIRM framework just presented, where the discounting distance parameter is 0, can be interpreted as a generalisation of the simple classical cost--loss decision model for dichotomous forecasts (e.g. \cite{richardson2003economic}). In this model, a user takes preventative action at cost $C$ if and only if the event is forecast. On the other hand, if the event is not forecast but occurs then the user incurs a loss $L$. It is assumed that $0<C<L$, otherwise the user would not take preventative action. This model can be encoded in an expense matrix $M_\mathrm{expense}$, whose $(i,j)$th entry is the expense incurred if $C_i$ is forecast and $C_j$ observed, namely
\[
M_\mathrm{expense} =
\begin{pmatrix}
0 & L \\
C & C
\end{pmatrix}.
\]
The expense matrix can be converted into a relative economic regret matrix $M_\mathrm{regret}$, the latter encoding the economic loss incurred relative to actions taken based on a perfect forecast. Explicitly,
\[
M_\mathrm{regret} =
\begin{pmatrix}
0 & L-C \\
C & 0
\end{pmatrix},
\]
where the $(i,j)$th entry gives the relative regret acting on the basis of forecast $C_i$ when $C_j$ was observed. For example, a miss (forecast $C_0$, observe $C_1$) incurs loss $L$, but even a perfect forecast (forecast and observe $C_1$) would incur cost $C$, so the relative economic regret is $L-C$. As noted by \cite{ehm2016quantiles}, from a decision theoretic perspective the distinction between expense and economic regret is inessential because the difference depends on the observations only. The matrix $M_\mathrm{regret}$ is precisely the dichotomous FIRM scoring matrix in Equation~(\ref{eq:2by2 and 3by3 scoring matrix}) for the choice $\alpha=1-C/L$ and $w_1=L$. Thus, over many forecast cases, the mean score is the average relative economic regret for a user whose decisions to take protective action were based on the forecast. The consistent forecast directive aligns with the well-known result that the user minimises their expected expense by taking protective action if and only if the probability of an event exceeds their \textit{cost--loss ratio} $C/L$.

To interpret the multicategorical case, the user takes a specific form of protective action at each threshold $\theta_i$ below the forecast category. The cost--loss ratio $C/L$ for each threshold is identical but the relative costs and losses differ by threshold $\theta_i$ according to the weights $w_i$. As before, the FIRM score is the economic regret relative to basing decisions on a perfect forecast.

\subsection{Example using NSW rainfall data}\label{ss:rainfall example}

To illustrate the FIRM framework, we use real forecasts of daily precipitation for 110 locations across New South Wales (NSW), Australia, for the two year period starting 1 April 2019. Two forecast systems of the BoM are compared: the Operational Consensus Forecast (OCF) and Official. OCF is an automated statistically post-processed poor man's ensemble \citep{bom2018upgrade}. Official is the official forecast published by the BoM and is manually curated by meteorologists. Both systems issue forecasts for the probability of precipitation exceeding various thresholds, from which we have reconstructed full predictive distributions using a hybrid generalised gamma distribution with very close fits to known points from the original distribution. The forecast data used here is from the reconstructed distributions.

Suppose that a tiered warning service for heavy rainfall has two category thresholds, $\theta_1=50\mathrm{mm}$ and $\theta_2=100\mathrm{mm}$, to demark three categories: `no warning conditions' ($C_0$), `heavy rainfall' ($C_1$) and `very heavy rainfall' ($C_2$). With specified weights $w_1=1$ and $w_2=4$ and risk parameter $\alpha=0.75$, the FIRM scoring matrix is
\begin{equation}\label{eq:rainfall matrix}
\begin{pmatrix}
	0			& 0.75		& 3.75 \\
	0.25 		& 0			& 3 \\
	1.25 		& 1 			& 0
\end{pmatrix},
\end{equation}
so that misses attract greater penalties than false alarms. This application is denoted by $\mathrm{FIRM}\big((50, 100), (1, 4), 0.75, 0\big)$.

A directive that is consistent with the optimal forecast strategy is to `Forecast the category that contains the 0.75-quantile of the predictive precipitation distribution.' Consequently, the optimal forecast strategy is to warn for very heavy rainfall at a location if the 0.75-quantile of the predictive distribution exceeds 100mm, or equivalently if the forecast probability of exceeding 100mm is greater than 25\%. 

Using this directive for converting a probabilistic forecast into a categorical forecast, contingency tables for lead day 1 Official and OCF forecasts are shown in Table~\ref{tab:rainfall contingency}. Note that the observed base rate in this sample for warning conditions ($C_1\cup C_2$) is about 4.5 times the observed base rate for very heavy rainfall events ($C_2)$, the latter being about 0.015\% (though base rates vary by location). The mean score for OCF was $7.2\times10^{-3}$ compared with $7.4\times10^{-3}$ for Official, indicating that OCF performed better overall. A 95\% confidence interval for the difference in the means is $(-1.3\times10^{-3}, 1.5\times10^{-3})$ and includes 0, which indicates that the difference in performance is not statistically significant at the 5\% level. See Appendix~\ref{a:CIs} for details and discussion on confidence interval estimation.

By writing the scoring matrix as a sum of its upper and lower triangular matrices, the mean score can also be expressed as a sum of the penalty from misses and from false alarms. This reveals stark differences between the two systems. The mean penalty for misses was $6.1\times10^{-3}$ (i.e., 87\% of the total mean score) for OCF in contrast to $3.6\times10^{-3}$ (49\% of the total) for Official. Conversely, Official was penalised heavily for false alarms relative to OCF.

\begin{table}[]
\begin{center}
\caption{Contingency table for NSW rainfall data, lead day 1.}
\label{tab:rainfall contingency}
\begin{small}
\begin{tabular}{cl|r|r|r|rcl|r|r|r|r}
\cline{3-5} \cline{9-11}
\multicolumn{1}{l}{\textbf{OCF:}}                        &                & \multicolumn{3}{c|}{\textbf{observed}} & \multicolumn{1}{c}{}                & \multicolumn{1}{l}{\textbf{Official:}}                  & \multicolumn{1}{c|}{} & \multicolumn{3}{c|}{\textbf{observed}} & \multicolumn{1}{c}{}                \\ \cline{3-6} \cline{9-12} 
\multicolumn{1}{l}{}                                     &                & $C_0$       & $C_1$       & $C_2$      & \multicolumn{1}{l|}{\textbf{total}} & \multicolumn{1}{l}{}                                    &                       & $C_0$       & $C_1$       & $C_2$      & \multicolumn{1}{l|}{\textbf{total}} \\ \hline
\multicolumn{1}{|c|}{\multirow{3}{*}{\textbf{forecast}}} & $C_0$          & 77984       & 259         & 37         & \multicolumn{1}{l|}{78280}          & \multicolumn{1}{c|}{\multirow{3}{*}{\textbf{forecast}}} & $C_0$                 & 77658       & 165         & 13         & \multicolumn{1}{l|}{77836}          \\ \cline{2-6} \cline{8-12} 
\multicolumn{1}{|c|}{}                                   & $C_1$          & 199         & 136         & 50         & \multicolumn{1}{l|}{385}            & \multicolumn{1}{c|}{}                                   & $C_1$                 & 451         & 171         & 36         & \multicolumn{1}{l|}{658}            \\ \cline{2-6} \cline{8-12} 
\multicolumn{1}{|c|}{}                                   & $C_2$          & 6           & 15          & 27         & \multicolumn{1}{l|}{48}             & \multicolumn{1}{c|}{}                                   & $C_2$                 & 80          & 74          & 65         & \multicolumn{1}{l|}{219}            \\ \hline
\multicolumn{1}{c|}{}                                    & \textbf{total} & 78189       & 410         & 114        & \multicolumn{1}{l|}{78713}          & \multicolumn{1}{c|}{}                                   & \textbf{total}        & 78189       & 410         & 114        & \multicolumn{1}{l|}{78713}          \\ \cline{2-6} \cline{8-12} 
\end{tabular}
\end{small}
\end{center}
\end{table}

\begin{figure}[bt]
\centering
\includegraphics{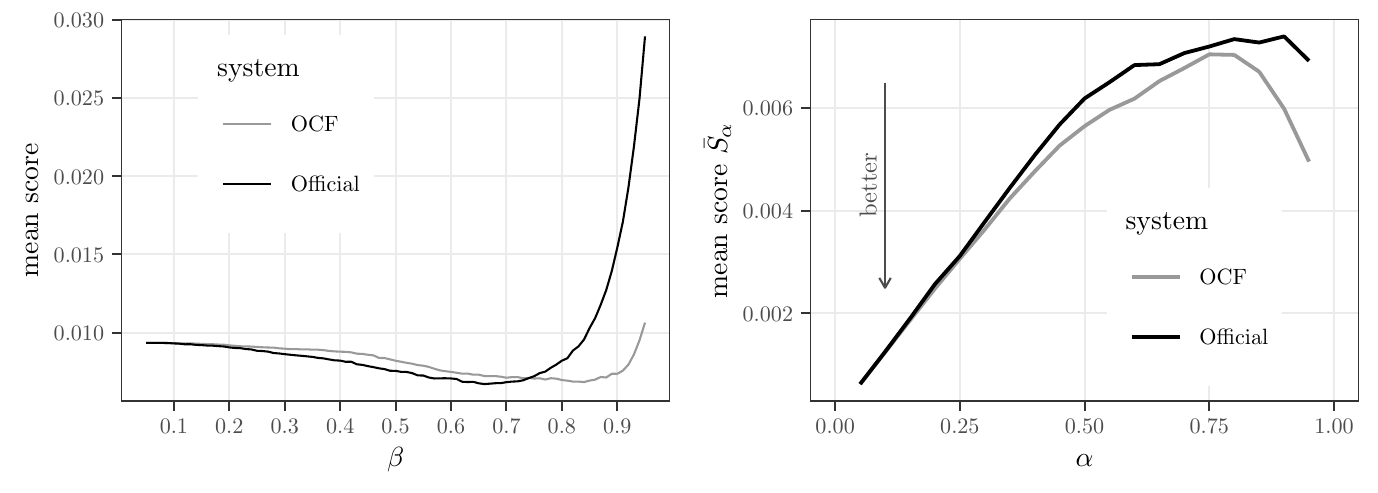}
\caption{Scores using NSW rainfall data. Left: Mean score using the FIRM scoring matrix with $\alpha=0.75$, but where the categorical forecast issued is one that contains a $\beta$-quantile of the predictive distribution, as discussed in Section~4.2. Right: Mean FIRM scores for different risk parameters $\alpha$, and where the categorical forecast issued is one that contains an $\alpha$-quantile of the predictive distribution, as discussed in Section~4.3.}
\label{fig:rainfall_multi}
\end{figure}

Neither OCF nor Official are perfectly calibrated. To understand the potential of both warning systems once calibrated, we examine their performance using the FIRM scoring matrix of Equation~(\ref{eq:rainfall matrix}), with $\alpha=0.75$, but where each system forecasts a category in which the $\beta$-quantile of their predictive distribution lies, for specified $\beta$. The results are shown in the left panel of Figure~\ref{fig:rainfall_multi} for a range of $\beta$ values. Official scored best when $\beta=0.66$. Calibrated Official ($\beta=0.66$) showed a 7\% improvement in score compared to uncalibrated Official ($\beta=0.75$). Using the method discussed in Appendix~\ref{a:CIs}, the null hypothesis that `Calibrated Official is no better than Official' can be rejected at the 5\% significance level. OCF scored best when $\beta=0.84$ with a 3\% improvement over the uncalibrated system. Hence Official has an over-prediction bias and OCF an under-prediction bias at the 50mm and 100mm category thresholds. For this sample, Calibrated Official would have performed marginally better than Calibrated OCF.

\subsection{Practical considerations when using the FIRM framework}\label{ss:practical}

We have introduced the FIRM framework for multicategorical forecasts. But those who use FIRM need to make appropriate choices for the parameters $\alpha$, $\theta_i$ and $w_i$. If the forecast service is intended for decision making in a specific commercial venture then the costs and losses in that operating environment should determine the parameters (c.f. \cite{ambuhl2010customer}). Consideration should be given to varying the parameters with forecast lead time, since forecast accuracy and the cost of taking protective action also vary with lead time \citep{jewson2021decide}.

For public weather warnings, the considerations are different (c.f. \cite{rothfusz2018facets}). Meteorology agencies have long histories of selecting (and sometimes revising) category thresholds $\theta_i$ for their warning services based on previous events, the impact of severe weather on communities, urban design regulations, consultation with emergency services and community engagement. Less often are risk parameters, such as $\alpha$, explicitly selected. Over-warning and false alarm intolerance can lead to warning fatigue, weaken trust in forecasts and willingness to respond appropriately to warnings \citep{gutter2018severe, potter2018influence, mackie2014warning, hoekstra2011preliminary}, and must be weighed against the cost of misses. Appropriate engagement with the community around risk tolerance, warning service design and communication is essential. Some recent studies attempt to quantify appropriate risk thresholds for public forecasts {\citep{rodwell2020user, roulston2004boy}, while insights from prospect theory are also informative \citep{kahneman1979prospect}. A study by \cite{leclerc2015cry} suggests that while false alarms can undermine trust in forecasts, this effect is only moderate compared with the stronger positive effect of including well-communicated probabilistic information with the forecast \citep{joslyn2009effects}. Thus $\alpha$ may be selected from a range of values and still be suitable for a public warning service, provided that the service is informed by best-practice warning communication and community engagement.

Appendix~\ref{a:choosing alpha} shows a method for estimating an unspecified yet implicit risk parameter $\alpha$ for an existing warning service, based on signal detection theory. Benchmarking the current service is useful for the process of establishing or modifying formal service definitions. This appendix also shows that the relationship between $\alpha$ and the proportion of misses to false alarms is not as straightforward as one might na\"ively expect, even for well-calibrated forecast systems. This has implications for how one describes the service to stakeholders.

For public warning services, we advise against using a different value of $\alpha$ for each categorical threshold $\theta_i$. Appendix~\ref{a:vary alpha} gives reasons. If there is a strong need in a public warning service to have a different risk parameter $\alpha$ for each categorical threshold $\theta_i$ , then one could consider implementing a separate warning product for each distinct categorical threshold rather than considering it as a single tiered warning service.

Design of public warning systems also includes the selection of an appropriate standard lead time \citep{hoekstra2011preliminary} and assessment on whether there is any benefit to warning early or late, noting that the urgency for people to pay attention may wane with increasing lead time \citep{turner1986waiting, mackie2014warning}. In situations where warnings are issued at more than one lead time, there can be compelling reasons for varying the risk parameter $\alpha$ with lead time. In many situations we anticipate that the threshold probability $1-\alpha$ for issuing a warning should be higher at the early warning lead time than at the standard lead time, to reduce the number of warnings that are retracted. Appendix~\ref{a:vary alpha with lead time} illustrates how one might choose suitable values for $\alpha$ at different lead times to achieve desired service outcomes.

As with all forecast verification, care must be taken when aggregating performance scores from multiple locations. Aggregated results are easiest to interpret when the risk parameter $\alpha$ is constant across the domain, and when the category thresholds $\theta_i$ vary across the domain so that the climatological base rates of each warning category are spatially invariant. The NSW rainfall example in Section~\ref{ss:rainfall example} used fixed thresholds 50mm and 100mm. Consequently, mean scores are higher in the wetter northeastern part of the domain than in the drier northwestern quarter, and hence the forecast system that performs best in the northeast is more likely to obtain a better mean score overall. 

To illustrate why $\alpha$ should be constant for meaningful aggregation, for a given $\alpha$ we calculate the mean score $\bar S_\alpha$ for the lead day 1 NSW precipitation forecasts, using the FIRM scoring matrix of Equation~(\ref{eq:framework scoring matrix}) for fixed $\theta_1=50$mm, $\theta_2=100$mm, $w_1=1$ and $w_2=4$. The category that is forecast in each case is the one which contains the $\alpha$-quantile of the predictive distribution. A graph of $\bar S_\alpha$ against $\alpha$ is shown in the right panel of Figure~\ref{fig:rainfall_multi}. For small $\alpha$, $\bar S_\alpha$ is low because the forecast problem is easy. Here, the $\alpha$-quantile is usually well below the 50mm threshold, so false alarms are rare while the misses are penalised very lightly.  For mid to high values of $\alpha$, $\bar S_\alpha$ is higher because the forecast problem is harder. There are more cases where the $\alpha$-quantile is near category thresholds, and so more chances of a false alarm or otherwise a more heavily penalised miss. This reemphasises that the FIRM scoring matrix of Equation~(\ref{eq:framework scoring matrix}) is not a normalised skill score, though a FIRM skill score can be constructed from it in the standard way \citep[p.~27]{potts2003basic}. Instead, it is designed as a consistent scoring function to monitor performance trends or to rank competing forecast systems that could be used for a multicategorical forecast service with specified threshold probability.

Finally, consideration should be given to the choice of weights $w_i$. If the frequency with which categories are observed is roughly equal and the consequences of a forecast error relative to one category threshold is no different from any other threshold, then a natural choice is $w_i=1$ for every $i$. However, for most tiered warning services, observations fall in higher categories less frequently yet the impact of forecast errors at these higher categories tends to be greater. In this context, applying equal weights will not appropriately reflect the costs of forecast errors relative to different category thresholds. Moreover, a forecast system that has good discrimination between events and nonevents relative to lower category thresholds but performs poorly for higher thresholds is unlikely to suffer a bad mean FIRM score over many events when equal weights are applied, because forecast cases that expose its weakness will be relatively rare. Instead, the weights should reflect the higher cost of poor discrimination at higher thresholds. If quantifying these costs is difficult, one simple approach is to calculate the base rate $r_i$ of observations exceeding $\theta_i$ over a fixed climatological reference period, and then set $w_i = 1/r_i$. If normalisation is desired, set $w_i = r_1/r_i$. The FIRM weights selected in the NSW rainfall example of Section~\ref{ss:rainfall example} loosely followed this principle.

\section{Extension of the FIRM framework}\label{s:variations}

\subsection{Reframing and proof of consistency}\label{ss:reframing and proof}

In Section~\ref{ss:basic framework} the basic FIRM framework for assessing ordered categorical forecasts was expressed in terms of a scoring matrix, so that forecasts could be scored using knowledge only of the forecast and observed categories. We now reframe the scoring method so that it is expressed in terms of the underlying continuous variables, namely some single-valued (or point) forecast $x$ and a corresponding observation $y$ in $I$. The two scoring formulations are equivalent, but this reframing allows an efficient proof that an optimal forecast strategy is consistent with the directive `Forecast a category that contains an $\alpha$-quantile of the predictive distribution.' It also facilitates notation for an extension of the framework presented thus far.

For $\theta$ in $I$ and $\alpha$ in $(0,1)$, let $S^\Qu_{\theta,\alpha}: I\times I\to [0,\infty)$ denote the scoring function
\begin{equation}\label{eq:S Q theta alpha}
S^\Qu_{\theta,\alpha}(x,y) =
\begin{cases}
1-\alpha, & y \leq \theta < x,\\
\alpha, & x \leq \theta < y,\\
0, & \text{otherwise.}
\end{cases}
\end{equation}
In the convention we have adopted, where higher values of $x$ and $y$ indicate more hazardous forecast or observed conditions, the scoring function $S^\Qu_{\theta,\alpha}$ applies a penalty of $1-\alpha$ for false alarms and a penalty of $\alpha$ for misses relative to the threshold $\theta$. So for a sequence $(\theta_i)_{i=1}^N$ of category thresholds and corresponding weights $w_i$, the scoring function $S^\Qu:I\times I\to [0,\infty)$, given by
\begin{equation}\label{eq: S Q}
S^\Qu(x,y) = \sum_{i=1}^N w_i\,S^\Qu_{\theta_i,\alpha}(x,y),
\end{equation}
is equivalent to the FIRM scoring matrix defined by Equation~(\ref{eq:framework scoring matrix}) via mapping the point forecast $x\in I$ and the observation $y\in I$ to the unique category in which each belongs.

The scoring function $S^\Qu_{\theta,\alpha}$ is an \textit{elementary scoring function} for the $\alpha$-quantile. Since $S^\Qu$ is a linear combination of the elementary scoring functions, it is \textit{consistent} for the $\alpha$-quantile \citep[Theorem 1]{ehm2016quantiles}. This means that any $\alpha$-quantile of the predictive distribution $F$ is a minimiser of the mapping $x\mapsto\EEE[S^\Qu(x,Y)]$, given that $Y$ has distribution $F$ \citep[Definition~2.1]{gneiting2011making}. Hence if $C_i$ contains an $\alpha$-quantile of $F$ then $C_i$ is a minimiser of the forecaster's expected FIRM score.

\subsection{A scoring function that discounts the penalty for marginal events}\label{ss:discount variation}

In many contexts it may be not be desirable to penalise forecast errors strictly categorically, where near and gross misses attract the same penalty \citep{barnes2007false, sharpe2016flexible}. The following variation on our categorical framework provides a more nuanced scoring system so that near misses and close false alarms are penalised less than gross misses and spectacular false alarms, whilst retaining the categorical nature of the forecast. The scoring method requires knowledge of the forecast category and of the real-valued observation.

Whenever $0<a\leq\infty$, $\theta\in I$ and $0<\alpha<1$, let $S^\Hu_{\theta,\alpha,a}: I\times I\to [0,\infty)$ denote the scoring function
\begin{equation}\label{eq:S H theta alpha}
S^\Hu_{\theta,\alpha, a}(x,y) =
\begin{cases}
(1-\alpha)\min(\theta - y, a), & y \leq \theta < x,\\
\alpha\min(y - \theta, a), & x \leq \theta < y,\\
0, & \text{otherwise,}
\end{cases}
\end{equation}
whenever $x,y\in I$. The parameter $a$ is called the \textit{discounting distance parameter}. When $a$ is finite, false alarms with respect to the threshold $\theta$ are typically penalised by $(1-\alpha)a$, but if the observation $y$ is within distance $a$ of the threshold $\theta$ then a discounted penalty is applied, being proportional to the distance of $y$ from $\theta$. Similar discounting occurs for misses that are within $a$ of the threshold $\theta$. When $a=\infty$, the cost of a miss is always proportional to the distance of the observation from the threshold, and similarly for false alarms. Note that the only information used about the point forecast $x$ is whether it lies above or below the categorical threshold $\theta$. Hence this scoring function can be written so that the forecast argument is categorical and the observation argument is real-valued.

To generalise this for multicategorical forecasts, we sum across all categorical thresholds $\theta_i$ to obtain the scoring function $S^\Hu$, where
\begin{equation}\label{eq: S H}
S^\Hu(x,y) = \sum_{i=1}^N w_i\,S^\Hu_{\theta_i,\alpha,a}(x,y)
\end{equation}
and, like Equation~(\ref{eq: S Q}), each positive weight $w_i$ specifies the relative importance of forecasting on the correct side of the categorical threshold $\theta_i$.

Given a predictive distribution $F$, one single-valued forecast $x$ in $I$ that optimises the expected score $S^\Hu$ is a so-called \textit{Huber quantile} $\Hu^\alpha_a(F)$ of the predictive distribution $F$ \citep[Theorem~5.2]{taggart2022point}.  Hence, when $S^\Hu$ is interpreted as a scoring function for categorical forecasts, it is consistent with the directive `Forecast any category $C_i$ that contains a Huber quantile $\Hu^\alpha_a(F)$.'

Huber quantiles are a type of generalised quantile \citep{bellini2014generalized} that can be traced back to the pioneering work of \cite{huber1964robust}. Like quantiles, a Huber quantile $\Hu^\alpha_a(F)$ for any given predictive distribution $F$ is not necessarily unique when $a$ is finite. However, for meteorological predictive distributions it is usually unique, and will be whenever the $\alpha$-quantile $\Qu^\alpha(F)$ is unique \citep{taggart2022point}. In the case when $a=\infty$, $\Hu^\alpha_a(F)$ is only defined if $F$ has finite first moment, in which case it will always be unique and is typically called the $\alpha$-expectile of $F$ \citep{newey1987asymmetric}. The special case $\Hu^{1/2}_\infty(F)$ is the well-known mean (or expected) value of $F$. An important property is that $\Hu^\alpha_a(F)\to \Qu^\alpha(F)$ as $a\downarrow0$, so that Huber quantiles are intermediaries between quantiles and expectiles. In particular, the Huber quantile $\Hu^{1/2}_a(F)$ is an intermediary between the median and mean values of $F$.

There are a number of ways of calculating the Huber functional or expectile of a predictive distribution $F$. One approach is to calculate the $\alpha$-quantile of a specific transformation of $F$ \citep{jones1994expectiles}. A different approach uses the fact that $x$ is a Huber quantile $\Hu^\alpha_a(F)$ if and only if it is a solution $x$ to the integral equation
\begin{equation}\label{eq:Huber functional geom interp}
\alpha \int_{[x,x+a]} (1-F(t))\,\dd t = (1-\alpha) \int_{[x-a,x]} F(t) \,\dd t
\end{equation}
\citep{taggart2022point}, and can thus be computed using numerical methods. Equation~(\ref{eq:Huber functional geom interp}) has a nice geometric interpretation in terms of the area above and below the graph of the predictive distribution $F$. The Huber quantile $\Hu^\alpha_a(F)$ is a point $x$ at which the ratio of the area below $F$ on the interval $[x-a,x]$ to the area above $F$ on the interval $[x,x+a]$ is $\alpha:(1-\alpha)$. 

\begin{figure}[bt]
\centering
\includegraphics{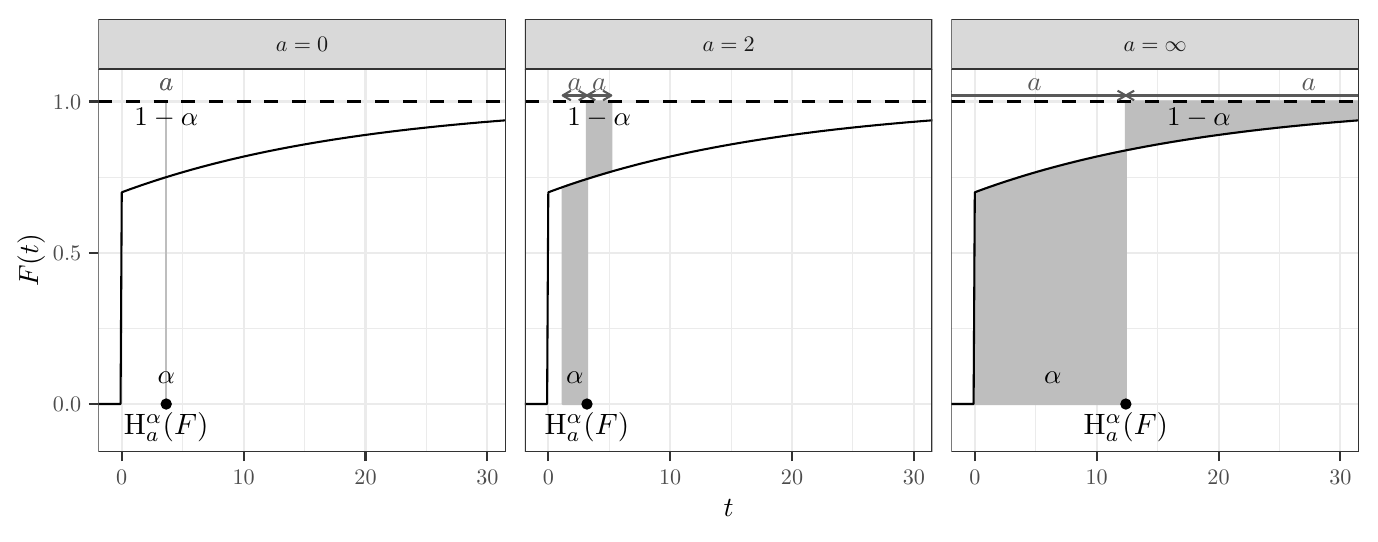}
\caption{Huber quantiles $\Hu^\alpha_a(F)$ for the distribution $F(t)=0.7 - 0.3\exp(-t/20)$, $t\geq0$, where $\alpha=0.75$. The Huber quantile $\Hu^\alpha_0(F)$ is identical with the $\alpha$-quantile while $\Hu^\alpha_\infty(F)$ is identical with the $\alpha$-expectile. In each case, the shaded regions below or above the graph of $F$ are of width $a$ and the ratio of the areas of the lower to upper regions is $\alpha:(1-\alpha)$.}
\label{fig:huber}
\end{figure}

This interpretation is illustrated in Figure~\ref{fig:huber} for the distribution $F(t)=0.7 - 0.3\exp(-t/20)$, $t\geq0$, which corresponds to a convective situation where the chance of precipitation is only 30\%, but if it does rain substantial falls are possible. The Huber quantile $\Hu^\alpha_2(F)$ when $\alpha=0.75$ is shown in the central panel, flanked by the limiting cases $a\downarrow0$ and $a\to\infty$ which correspond to the $\alpha$-quantile and $\alpha$-expectile respectively. The quantile $\Hu^\alpha_0(F)$, Huber quantile $\Hu^\alpha_2(F)$ and expectile $\Hu^\alpha_\infty(F)$ are all risk measures which could be used to prompt a warning if they exceed some specified categorical warning threshold $\theta$. The figure shows that the quantile ignores information in the tail of the predictive distribution, while the expectile uses that information and hence for this particular distribution is greater than the quantile. It is in the tail where the extremes typically lie. For this reason, along with a number of other properties, expectile forecasts have recently attracted interest in financial risk \citep{ehm2016quantiles, bellini2017risk}. As a risk measure, the Huber quantile is a compromise between the expectile and quantile.

While applying a discounted penalty for near misses and close false alarms is an attractive option, there are downsides. The scoring method cannot be written as scoring matrix. Instead of maintaining a contingency table, one must keep track of forecast categories and corresponding real-valued observations. Secondly, Huber quantiles and expectiles will be unfamiliar to most people, whereas quantiles (or percentiles) are more widely known. Nonetheless, a compelling reason to use a scoring function like $S^\Hu$ instead of $S^\Qu$ is that in many situations it provides a better model of the economic costs of forecast errors than the classical cost--loss model \citep{ehm2016quantiles, taggart2022point}.

\subsection{Unification of the framework}\label{ss:unification}

The scoring functions $S^\Qu$ and $S^\Hu$ of Equations~(\ref{eq: S Q}) and (\ref{eq: S H}) describe what may appear to be two different families of scoring functions. However, the pointwise limit $\frac{1}{a}S^\Hu(x,y) \to S^\Qu(x,y)$ as $a\downarrow0$ for fixed $\theta_i$, $w_i$ and $\alpha$ \citep{taggart2022point} shows that mathematically they belong to the same unified family. Hence any particular application of the FIxed Risk Multicategory framework presented in this paper can be summarised using the notation
\[
\mathrm{FIRM}\left((\theta_i)_{i=1}^N, (w_i)_{i=1}^N, \alpha, a\right),
\]
where $(\theta_i)_{i=1}^N$ denotes the chosen categorical thresholds, $(w_i)_{i=1}^N$ the corresponding scoring weights, $\alpha$ the risk parameter (so that the cost of a miss relative to a false alarm is in the ratio $\alpha:(1-\alpha)$) and $a$ the discounting distance parameter. In the case when $a=0$, the corresponding consistent scoring function is given by Equation~(\ref{eq: S Q}), which is equivalent to the scoring matrix defined by Equation~(\ref{eq:framework scoring matrix}), and the consistent forecast directive is `Forecast any category that contains an $\alpha$-quantile of the predictive distribution.' In the case when $0<a\leq\infty$, the consistent scoring function is given by Equation~(\ref{eq: S H}) and the forecast directive is `Forecast any category that contains a Huber quantile $\Hu^\alpha_a(F)$ of the predictive distribution $F$.' For the special case when $a=\infty$ and $\alpha=1/2$, the consistent forecast directive is `Forecast the category that contains the mean value of the predictive distribution.' The specified weights $w_i$ do not affect the forecast directive nor the optimal conversion rule from a predictive distribution to a categorical forecast. Instead, they indicate the category thresholds for which discrimination of events from nonevents is most valuable by providing commensurate penalties for poor discrimination, and in this sense guide where to focus improvements in forecast system development.

\subsection{Categorical forecasts for the likelihood of an event}\label{ss:binary variation}

A special case of the FIRM framework covers the situation where the observations are dichotomous (event or nonevent) and the forecast categories indicate the likelihood of the event. Let $y$ denote the observed outcome, where $y=1$ if the event occurs and $y=0$ if not. The forecast probability $p$ of event occurring is precisely the mean of the predictive distribution $F$, with $p$ taking values in the prediction space $I=[0,1]$. So in the FIRM framework context, $a=\infty$, $\alpha=1/2$, $x=p=\Hu^{1/2}_\infty(F)$ and the forecast categories $C_i$ are determined by the category thresholds $\theta_i$ of likelihood.

For example, suppose that the forecast categories are a `low,' `medium' and `high' chance of lightning. If the forecast directive is ``Forecast `low' when $\PPP(\mathrm{lightning})\leq 0.33$, `high' when $\PPP(\mathrm{lightning})>0.67$ and `medium' otherwise,'' then the thresholds are $\theta_1=0.33$ and $\theta_2=0.67$. Although the underlying probability forecast could be assessed directly, the quality of the categorical forecasts is better assessed by ascertaining how well a forecast system discriminates events from nonevents at the specified category thresholds $\theta_i$. Below we produce a scoring matrix that does that, with weights $w_i$ selected according to the relative importance of discrimination at each $\theta_i$.

By results of Section~\ref{ss:discount variation} and the fact that $x=p=\Hu^{1/2}_\infty(F)$, the scoring function $S^\Hu$ of Equation~(\ref{eq: S H}), with $\alpha=1/2$ and $a=\infty$, is consistent with the directive `Forecast the category which contains the forecast probability $p$ of the event'. In this binary setting, $S^\Hu$ can be converted into a scoring matrix by rewriting the elementary scoring function $S^\Hu_{\theta,\alpha,a}$ of Equation~(\ref{eq:S H theta alpha}) as
\[
S^\Bi_\theta(p,y) = 2\,S^\Hu_{\theta,1/2,\infty}(p,y) = 
\begin{cases}
\theta, & y=0, \, p > \theta \\
1-\theta, & y=1,\,  p \leq \theta, \\
0, & \mathrm{otherwise}.
\end{cases}
\]
Then we obtain the consistent scoring function
\begin{equation}\label{eq: S B}
S^\Bi(p,y) = \sum_{i=1}^N w_i\,S^\Bi_{\theta_i}(p,y),
\end{equation}
which is equivalent to $S^\Hu$ via a rescaling of weights $w_i$. Using the natural mapping from $p$ to the category $C_i$ that contains $p$, the scoring function $S^\Bi$ can be written as a scoring matrix whose $(i,j)$th entry is the penalty applied when $C_i$ is forecast and the outcome is $j$. In the 3-category case, the scoring matrix is
\[
\begin{pmatrix}
	0							& w_1(1-\theta_1) + w_2(1-\theta_2) \\
	w_1 \theta_1 		& w_2(1-\theta_2) \\
	w_1 \theta_1 + w_2 \theta_2 &  0
\end{pmatrix},
\]
where the left column corresponds to an observed nonevent and the right column to an observed event. Nonzero scores in the left column correspond to penalties for false alarms, and nonzero scores in the right column to penalties for misses, thereby yielding a useful decomposition of the score.

\section{Summary and conclusion}

We have introduced a flexible family of scoring functions for ordered multicategorical forecasts, which are consistent with forecast directives based on fixed risk measures (quantiles, Huber quantiles and expectiles). To apply the FIxed Risk Multicategorical (FIRM) framework to a particular multicategorical forecast service, one needs to specify (i) categorical thresholds $\theta_i$, (ii) corresponding weights $w_i$, (iii) a risk parameter $\alpha$ and (iv) a discounting distance parameter $a$. Once specified, the framework gives a consistent scoring function and forecast directive. The scoring methodology is relatively easy to explain to stakeholders. For example, if $\alpha=0.75$ then the cost of a miss relative to a false alarm is the ratio 0.75:0.25 (equivalently 3:1). If $(w_1,w_2) = (1, 4)$ then forecast errors with respect to the second categorical threshold will be penalised 4 times more heavily than forecast errors with respect to the first categorical threshold. If $a=5$ then any forecast error where the real-valued observation is within 5 units of the forecast category receives a reduced penalty.

This flexibility in parameter choice allows the designers of multicategorical forecast services or tiered warning services to choose parameters that are suitable for the use case. Meteorological services typically have experience in specifying category thresholds $\theta_i$ (such as 34 knots being the lower threshold for a marine gale warning), and often there has been a desire to employ discounted penalties for marginal events, although until now such discounting had not been presented in a decision-theoretic coherent framework. We have indicated some factors that may be considered when selecting weights $w_i$ and the risk parameter $\alpha$, and presented a method for estimating the implicit risk level of an existing service where $\alpha$ has not been made explicit.

The FIRM framework for scoring multicategorical forecasts contrasts with many alternative performance measures that are currently recommended in the literature. In Section~\ref{s:critique} it was shown that some of the most popular measures are constructed based on assumptions that are often not aligned with sensible forecast service objectives. 

We conclude with a plea that the designers of categorical forecast services, including warning services, give careful thought to service definitions and how performance is assessed. An automated warning product trained to optimise the chosen performance measure may outperform its competitors, but if the performance measure does not reward forecasts with desired service properties then it may provide poorer outcomes for users. Only careful service design, where the chosen performance measure is aligned with user benefits, can give confidence that forecast verification is a reliable guide to service improvement.

\hspace{4mm}

\noindent\textit{Acknowledgements.} The authors wish to thank Brenda Mackie with whom they had very helpful conversations about the relevant social science literature, and Beth Ebert, Ben Hague and two anonymous reviewers for their feedback on earlier versions of this manuscript. The first author thanks his family for their patience and encouragement whilst writing this paper during lockdown.

\appendix

\section{Appendices}

\subsection{Synthetic data}\label{a:synthetic}

Appendices~\ref{a:pod and far} and \ref{a:choosing alpha} use the following synthetic data in a dichotomous setting. Suppose that a random variable $Y$ has a normal distribution with mean $\mu$ and standard deviation $\sigma$, i.e., $Y\sim\cN(\mu,\sigma^2)$. Let the category $C_1$ have a base rate (i.e. climatological relative frequency) of $r_1$. This implies that the single threshold $\theta_1$ at the boundary of the two categories satisfies $r_1=\PPP(Y>\theta_1)$ and so $\theta_1 = \mu+\sigma\Phi^{-1}(1-r_1)$.

We construct different forecast systems that issue perfectly calibrated predictive distributions $F$ for $Y$, but with varying degrees of sharpness (predictive precision) specified by a positive variable $\sigma_2$, where a smaller value of $\sigma_2$ indicates a more accurate system. For given $\sigma_2$, the forecast system is constructed as follows. The variable $Y$ is written as a sum $Y=Y_1+Y_2$ of two independent random variables satisfying $Y_1\sim\cN(\mu,\sigma_1^2)$, $Y_2\sim\cN(0,\sigma_2^2)$ and $\sigma^2=\sigma_1^2+\sigma_2^2$. The forecast system has knowledge of $Y_1$ and issues the perfectly calibrated predictive distribution $F$, where $F\sim Y_1+\cN(0,\sigma_2^2)$. We call $\sigma_2/\sigma$ the \textit{relative predictive uncertainty} of the system, since a value of 0 indicates perfect knowledge of $Y$ while a value of 1 indicates predictive skill identical to that of a climatological forecast.

In each example where this set-up is used, the category threshold $\theta_1$ is uniquely determined by the base rate $r_1$ and each forecast system is identified with its relative predictive uncertainty. In this way, one can obtain results applicable to a wide range of idealised observational distributions and corresponding forecast systems, independent of the specific choice of $\mu$ and $\sigma$. For ease of reference, each example uses four different base rates (0.01, 0.05, 0.1 and 0.25) and four systems with different relative predictive uncertainties (0.01, 0.1, 0.25, 0.5).

\subsection{POD and FAR synthetic experiment}\label{a:pod and far}

Suppose that a forecast system issues perfectly calibrated predictive distributions that are normal distributions, as per the set up of Appendix~\ref{a:synthetic}, and that the performance target to be met is $\mathrm{POD} \geq 0.7$ and $\mathrm{FAR} \leq 0.4$. One strategy is to warn if and only if the forecast probability $p_1$ of an event exceeds a threshold probability $1-\alpha$, where $\alpha$ is chosen to maximise the expected likelihood of meeting the target. 

Figure \ref{fig:pod far targets} shows accurate estimates for the expected likelihood of meeting the target for different values of $\alpha$, base rate, and predictive sharpness (as measured by relative predictive uncertainty). These are calculated using a standard bootstrapping method, with bootstrap sampling size sufficiently large so that standard errors for each estimate are less than 0.008.

\begin{figure}[bt]
\centering
\includegraphics{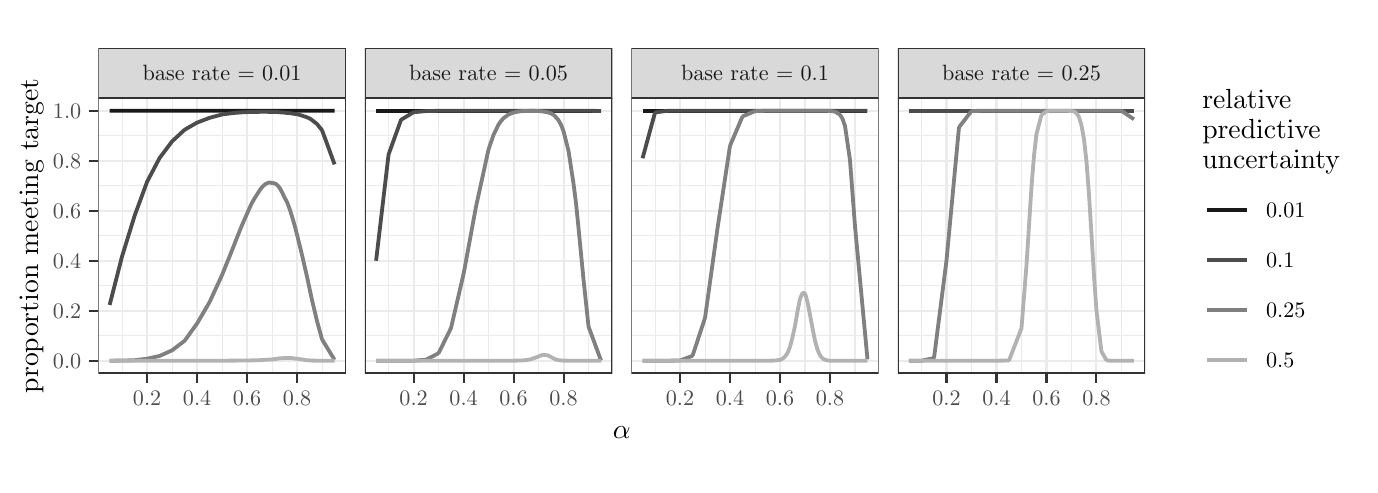}
\caption{The proportion of independently generated forecasts that meet the target of $\mathrm{POD} \geq 0.7$ and $\mathrm{FAR} \leq 0.4$. The underlying data and forecast systems are described in Appendix~A.2. A system warns if and only if the forecast probability of an event exceeds $1-\alpha$.}
\label{fig:pod far targets}
\end{figure}

\subsection{Confidence intervals and hypothesis testing}\label{a:CIs}

Section~\ref{ss:rainfall example} asserted that (a) there was no statistical significance between the predictive performance of OCF and Official at the 5\% significance level, and (b) Calibrated Official performed better than Official at the 5\% level. These inferences were based on the confidence interval (CI) estimates for the difference in mean FIRM scores as follows.

First, we removed spatial dependence by calculating the mean FIRM score for each date, which results in a time series of mean daily FIRM scores of length 731 for each forecast system. The resulting times series of differences of mean daily scores are zero inflated and not independent. For example, the difference in daily mean scores between OCF and Official has a lag-1 Pearson's correlation coefficient of 0.34, and is nonzero on only 32 out of 731 days.

The desired CI is estimated using a modification by \cite{harvey1997testing} of the Diebold--Mariano (DB) test statistic \citep{diebold1995comparing}. The method allows for serially correlated data. We treat the data as scores for 2-step-ahead forecasts, since lead day 1 forecasts are issued approximately 35 hours ahead of the realisation. Subject to traditional regularity conditions, the modified DM test statistic asymptotically has a Student's $t$-distribution, though convergence may be slow when derived from populations with zero inflated data.

\begin{table}[]
\begin{center}
\caption{95\% CIs estimated using different methods for the difference $\mu_\rA-\mu_\rB$ in the means of the daily mean FIRM scores for competing forecast systems A and B.}
\begin{small}
\label{tab:CIs}
\begin{tabular}{|l|l|r|l|r|r|}
\hline
Forecast A & Forecast B & $\mu_\rA-\mu_\rB$ \ ($\times10^4$) & Method & 95\% CI \ ($\times10^{4}$) \\ \hline
Official	& OCF 						& $1.46$ 	& Student's $t$	& $(-9.35, 12.26)$ \\ 
           	&								&				& ELR				& $(-11.41,12.16)$ \\
 		  	&                  			&				& DM					& $(-12.59, 15.50)$ \\ 
 		  	& 								&				& bootstrap 		& $(-13.29, 13.59)$ \\ \hline
Official 	& Calibrated Official	& $5.04$	& Student's $t$	& $(-1.14, 11.23)$ \\ 
           	&                     			& 				& ELR        		& $(-2.34, 11.17)$ \\
 		  	&                  			&				& DM					& $(-0.36, 10.45)$ \\ 
           	&                     			& 				& bootstrap 		& $(0.16, 10.79)$ \\ \hline 
\end{tabular}
\end{small}
\end{center}
\end{table}

Estimated 95\% CIs using the DM method are compared with estimates using three other methods in Table~\ref{tab:CIs}. The commonly used Student's $t$-distribution approximation (see, e.g., \cite{gilleland2020bootstrap}, p.~2123) is not recommended in this application since the assumption of serial independence is violated here, and it offers no advantage over the DM method in relation to zero inflated data.

CIs can be estimated using the non-parametric empirical likelihood ratio (ELR) statistic, which asymptotically has a chi-squared distribution \citep{owen1988empirical}. \cite{chen2003empirical} show that for a variety of populations with zero inflated data, the ELR method gives CI estimates with better properties when compared with estimates based on the $t$-distribution. However, the ELR method also assumes serial independence.

Bootstrapping (see, e.g., \cite{gilleland2020bootstrap}) provides an alternative. Circular block bootstrap sampling provides a way to respect serial correlation in the data. Here we used block sizes of length 27 (which is approximately $\sqrt{731}$, and ``much longer than the length of dependence, but much shorter than the entire series'' \citep{gilleland2020bootstrap}), generated 27,000 bootstrap samples and applied the \textit{percentile bootstrap method} \citep[p.~2125]{gilleland2020bootstrap} to estimate the CIs. The limitation is that bootstrapped samples are drawn from data with a small proportion of nonzero data points. 

On balance, the authors feel that the modified DM method or bootstrapping are the best of the options considered here. Assertions (a) and (b), stated at the beginning of this appendix, can be inferred from the estimated CIs of both methods, noting that (b) is based on one-sided CIs not shown in Table~\ref{tab:CIs}.

The challenge that serially correlated zero inflated data poses for estimating CIs for the difference in mean scores is not unique to FIRM scores. It also occurs when evaluating forecasts for extreme or rare events using threshold-weighted consistent scoring functions \citep{taggart2021evaluation}. The authors welcome further research on how to conduct statistical inference in this context.

\subsection{Estimating an unspecified risk parameter $\alpha$}\label{a:choosing alpha}

Suppose that an existing warning service does not explicitly specify a confidence threshold $1-\alpha$ for issuing a warning. It is possible to estimate an implicit confidence threshold from historical contingency tables. A na\"ive  approach assumes that the historical ratio of false alarms to misses indicates the implicit ratio of the cost of a miss to that of a false alarm. That is, an estimate $\hat\alpha$ of $\alpha$ is based on the equation $\hat\alpha m = (1-\hat\alpha) f$, which rearranged gives $\hat\alpha = f/(f+m)$.
However, $\hat\alpha$ is quite a biased estimator of $\alpha$. Using the synthetic data set-up of Appendix~\ref{a:synthetic}, 
we calculate $\hat\alpha$ as a function of $\alpha$ from $2\times10^7$ forecast cases. The results are plotted in the top panel of Figure~\ref{fig:alphahat}. The estimate is particularly bad when $\alpha$, forecast accuracy and base rates are low.

\begin{figure}[bt]
\centering
\includegraphics{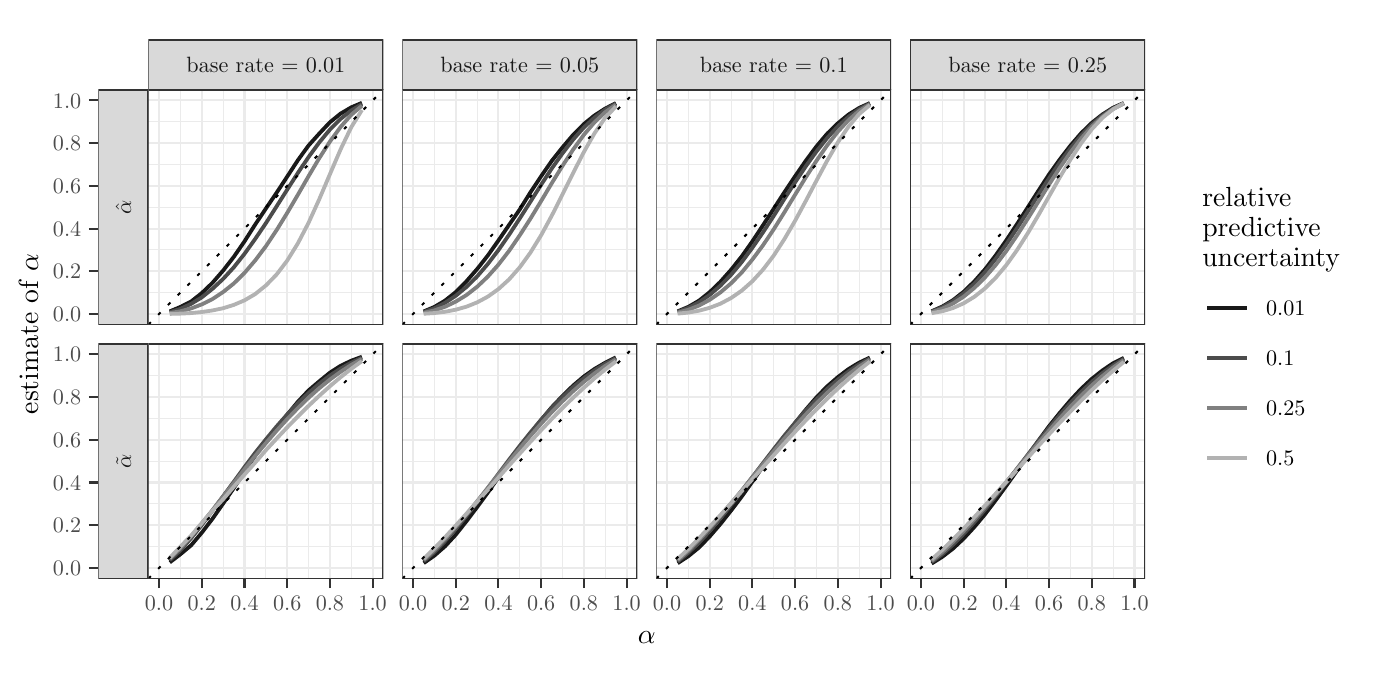}
\caption{The empirical relationship between $\alpha$ and estimators $\hat\alpha$ (top panel) and $\tilde\alpha$ (bottom panel) for perfectly calibrated forecast systems, as described in Appendix~A.1, based on $2\times10^7$ forecast cases.}
\label{fig:alphahat}
\end{figure}

Precise analytical statements can also be made. For example, a perfectly calibrated forecast system that warns if and only if the forecast probability of an event exceeds 0.5 is not expected to produce misses and false alarms in equal measure. To see why, assume again the synthetic data set-up, with random variable $Y$ satisfying $Y = Y_1+Y_2$ and perfectly calibrated predictive distributions of the form $F = Y_1 + \cN(0,\sigma_2^2)$. If the warning threshold is $\theta$ then the system warns if and only if $Y_1 > \theta$, since $\Qu^{0.5}(F) = Y_1$. Now
\begin{align*}
\PPP(\text{event observed}) &= \PPP(Y>\theta) = 1 - \Phi\left(\frac{\theta - \mu}{\sigma_1^2 + \sigma_2^2}\right), \\
\intertext{while}
\PPP(\text{warning issued}) &= \PPP(Y_1>\theta) = 1 - \Phi\left(\frac{\theta - \mu}{\sigma_1^2}\right).
\end{align*}
So if $\theta>\mu$ then the probability of issuing a warning is less than the probability of observing the event, and one can expect that there will be more misses than false alarms notwithstanding perfect calibration. 

A corollary is that, for well-calibrated forecast systems, the relative number of misses to false alarms is not necessarily a good indicator of performance. Conversely, calibrating a system so that $\alpha m \approx (1-\alpha) f $ may result in a poorer warning service as assessed by the consistent FIRM scoring matrix. When communicating to users, one should be hesitant in making statements that relate $\alpha$ to the proportion of false alarms to misses.

A different approach to estimating $\alpha$ from historical contingency tables comes from signal detection theory (e.g. \cite{mason2003binary}). If one assumes that the theory's `noise' and `signal plus noise' distributions are Gaussian with equal variance, then one obtains an estimate $\tilde\alpha$ of $\alpha$ given by
\begin{equation}\label{eq:alpha STD estimate}
\tilde\alpha = \frac{1}{\tau +1}, \qquad \tau = \frac{\phi(\Phi^{-1}(1-\mathrm{POD}))}{\phi(\Phi^{-1}(1-\mathrm{POFD}))}\times\frac{h+m}{f+c}.
\end{equation}
\citep[Section 3.4.4c]{mason2003binary}, where the probability of false detection (POFD) is given by $\mathrm{POFD} = f/(f+c)$. Figure~\ref{fig:alphahat} shows that for perfectly calibrated normal predictive distributions, $\tilde\alpha$ is generally a more reliable estimate of $\alpha$ than $\hat\alpha$, particularly for low base rates and less accurate forecast systems.

To illustrate, we convert the 3-category contingency tables of Table~\ref{tab:rainfall contingency} into two dichotomous contingency tables by merging $C_1$ and $C_2$ into a single category $C_1\cup C_2$. Then $\tilde\alpha=0.75$ for OCF and $\tilde\alpha=0.89$ for Official. However, neither forecast system is perfectly calibrated, nor is $\tilde\alpha$ an unbiased estimator of $\alpha$. With a base rate of $0.007$, the bottom left panel of Figure~\ref{fig:alphahat} suggests that these values of $\tilde\alpha$ may over-estimate $\alpha$ by around 0.05 or 0.1. Applying this correction gives estimated $\alpha$ values that are compatible with the earlier observation that the $0.75$-quantile of OCF had an under-forecast bias while that for Official had an over-forecast bias.

\cite{schmidt2021interpretation} present a method for identifying the statistical functional associated with an unknown forecast directive, based on time series of real-valued forecasts and observations. This method is free of the kind of assumptions used to justify Equation~(\ref{eq:alpha STD estimate}), and hence an extension of their method to ordered categorical forecasts would be welcome.

\subsection{Varying $\alpha$ with threshold}\label{a:vary alpha}

One could modify the FIRM framework by choosing different risk parameter $\alpha_i$ for each categorical threshold $\theta_i$.  For example, one could replace the scoring function $S$ of Equation~(\ref{eq: S Q}) with
\[
S(x,y) = \sum_{i=1}^N w_i\,S^\Qu_{\theta_i,\alpha_i}(x,y).
\]
However, we argue that this should only be done if it actually reflects the costs of forecast errors for the specific user. For public warning services, there are at least two reasons to avoid varying $\alpha$ with threshold. First, the forecast directive is considerably more complex than `Forecast a category that contains an $\alpha$-quantile of the predictive distribution'. Second, normally when there is a knife-edge decision as to which category to forecast, we would expect it to be a choice between adjacent categories. This is true when $\alpha$ is fixed but can be violated when $\alpha$ varies.

To illustrate this second point, consider a three-tiered warning service with $\theta_1 = 0$, $\theta_2 = 2$, $\alpha_1=0.1$, $\alpha_2=0.9$ and $w_1=w_2=1$. Suppose that the predictive distribution $F$ is $\cN(1,1)$, so that $p_0=0.16$, $p_1=0.68$ and $p_2=0.16$. Which warning category minimises the expected score? A quick calculation shows that, in general,
\[
\EEE[S^\Qu_{\theta,\alpha}(x,Y)]
=
\begin{cases}
(1-\alpha)F(\theta), & x>\theta,\\
\alpha(1-F(\theta)), & x\leq\theta,
\end{cases}
\]
whenever $Y$ has predictive distribution $F$. Consequently, when $Y\sim\cN(1,1)$ it can be shown that the forecaster's expected score $\EEE[S(x,Y)]$ is minimised whenever $x$ lies in $C_0$ or $C_2$ but not in $C_1$. Hence forecasting either $C_0$ or $C_2$ is optimal whilst forecasting $C_1$ is suboptimal.

\subsection{Risk parameter $\alpha$ varying with lead time}\label{a:vary alpha with lead time}

The following scenario illustrates why the risk parameter might vary with lead time. Consider a dichotomous warning service with two issue lead times (`early' and `standard'), and suppose that warnings are issued at standard lead time if and only if the probability of an event exceeds $1-\alpha$. Those designing the service specify that (i) it is slightly undesirable to not warn early then warn at standard lead time, and (ii) highly undesirable to warn early then retract the warning at the standard lead time. That is, warning early has some benefit but needs to be weighed against the heavy reputational cost of retracting warnings. This could be quantified by a penalty matrix $T$, whose $(i,j)$th entry $t_{ij}$ specifies the cost of forecasting category $C_i$ early and then forecasting $C_j$ at the standard lead time. To calculate the suitable threshold probability $1-\beta$ for issuing an early warning, an historical forecast data set can be used to find a $\beta$ that minimises the score
\[
S(\beta, \alpha) = \sum_{i=0}^1\sum_{j=0}^1 t_{ij}n_{ij}^{(\beta,\alpha)},
\]
where $n_{ij}^{(\beta, \alpha)}$ is the number of times that $C_i$ was forecast early based on the available $\beta$-quantile forecast, and that $C_j$ was forecast at standard time based on the available $\alpha$-quantile forecast.

\begin{figure}[bt]
\centering
\includegraphics{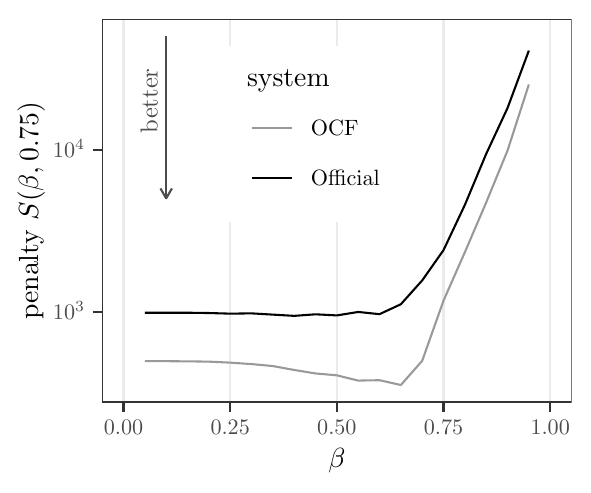}
\caption{Penalty when issuing warnings at lead day 2 based on $\beta$-quantile forecasts, given that lead day 1 warnings will be issued using $0.75$-quantile forecasts, as discussed in Appendix~A.6. The vertical scale is logarithmic.}
\label{fig:penalty}
\end{figure}

To illustrate this for the NSW rainfall data of Section~\ref{ss:rainfall example}, we take the categorical warning threshold $\theta_1=50\mathrm{mm}$, and suppose that the risk parameter $\alpha=0.75$ at the standard issue time (lead day 1) has been set. Suppose that retracting an early warning should be penalised 15 times greater than issuing a warning at standard lead time only. Then the penalty matrix $T$ is given by
\begin{equation}\label{eq:early warning matrix}
T = 
\begin{pmatrix}
	0			& 1 \\
	15 		& 0
\end{pmatrix}.
\end{equation}
The score $S(\beta, \alpha)$ is then calculated for a range of $\beta$ values associated with early warning decisions (lead day 2). Figure~\ref{fig:penalty} shows the results. On this dataset, $\beta=0.65$ was best for OCF and $\beta=0.6$ for Official. Unsurprisingly, warning early requires higher confidence than warning at the standard time. OCF scores better because it exhibits more stability across lead time.


\end{document}